\documentclass[preprint,12pt]{elsarticle}
\usepackage{amsmath}
\usepackage{amssymb}
\usepackage[pdftex,colorlinks]{hyperref}
\usepackage{multirow}
\usepackage{enumerate}
\usepackage{cases}
\usepackage{graphicx}
\usepackage{subfigure}
\usepackage{float}
\usepackage{mathrsfs}%花写体
\usepackage{multirow}%表格
\usepackage{booktabs}

\makeatletter

\newcommand {\Rmnum} [1] {\expandafter \@slowromancap \romannumeral #1@}
\makeatother
\newtheorem{Definition}{Definition}
\newtheorem{Protocol}{Protocol}

\biboptions{compress}

\begin{document}

\begin{frontmatter}

\title{Practical quantum oblivious transfer and bit commitment protocols}


%\author[label1]{Li Yang\corref{1}}\ead{yang@is.ac.cn}
%\cortext[1]{Corresponding author.}
%\ead{yangli@iie.ac.cn}

\author{Ya-Qi Song $^{1,2,3}$}
\author{Li Yang$^{1,2}$\corref{1}}%\ead{yangli@iie.ac.cn}
\cortext[1]{Corresponding author. Email: yangli@iie.ac.cn}
\address{1.State Key Laboratory of Information Security, Institute of Information Engineering, Chinese Academy of Sciences, Beijing 100093, China\\
2.Data Assurance and Communication Security Research Center,Chinese Academy of Sciences, Beijing 100093, China\\
3.University of Chinese Academy of Sciences, Beijing, 100049, China}

\begin{abstract}
We propose a practical quantum oblivious transfer and a bit commitment protocols which replace the single-photon source with weak coherent pulses and allow error and loss in channel and detectors. These protocols can be realized with available optoelectronic apparatus.
\end{abstract}


\end{frontmatter}

% \linenumbers

\section{Introduction}
Quantum oblivious transfer(QOT) and quantum bit commitment(QBC) protocols are basic aspects of quantum cryptography. The study of QOT was started by Cr\'{e}peau and Kilian \cite{1stQOT88}. In 1992, a practical QOT protocol was been proposed \cite{OT92}. However, in these two protocols, if Bob measures the pulses after Alice discloses the bases, he will get both messages and Alice's privacy can be destroyed. In 1993, a well-known QBC scheme was been presented \cite{BCJL}, which usually referred to as BCJL scheme was once believed as a provably secure scheme. Cr\'{e}peau proposed a QOT protocol \cite{Cre94} on top of this QBC scheme in 1994 to ensure Bob cannot delay his measurement. Then Yao proved it a secure protocol based on the security of QBC \cite{Yao95}. Unfortunately, Mayers found that the BCJL scheme was insecure \cite{Mayers96}. Later, Mayers, Lo and Chau separately present no-go theorem and prove that there is no non-interactive quantum bit commitment protocols with statistical security \cite{Mayers97,Lo97,BCMno-go97}. Therefore, researchers believe that QOT protocols whose security are usually based on QBC are insecure.

However, it isn't the only way to construct a QOT protocol based on QBC. Several researchers have made efforts in this direction\cite{HeOT06,QKDOT06,PUFOT10}. In this paper, we construct a random oblivious transfer protocol based on non-orthogonal states transmission over a quantum channel. Then, as that usually do in modern cryptography\cite{ROT=12}, construct a one-out-of-two oblivious transfer protocol on top of the R-OT protocol. In this scheme, Alice does not tell Bob the bases and there is no need to forbid Bob to delay the measurement by QBC protocols\cite{Yang13}. The protocols can be applied in the practical application allowing the imperfect of sources, the quantum channel and detectors. Then we present a bit commitment protocol based on this QOT protocol. Considering error-correcting code and tolerable error rate, we describe the protocols in detail and analyze the security and problem we probably face to in practice.

\section{Conditions in practice}
As a practical protocol, these situations should be considered:
\begin{enumerate}
\item Alice's emission apparatus. As practical and efficient single-photon sources have yet to be realized, weak coherent pulses with typical average photon number of $\mu_{S}$ which can be easily prepared by standard semiconductor lasers and calibrated attenuators are used in the following protocols. The error rate caused by the emission apparatus is $\varepsilon_{S}$. A pulse is requested to contain only one kind of polarization but more than one photons in a pulse are allowed.
\item Channel loss and error. The existence of channel loss leads to an imperfect transfer efficiency. And the noises in channel lead to channel error. Suppose the transfer efficiency of the channel is $\eta_C$, the error rate caused by channel is $\varepsilon_{C}$.
\item Bob's detection apparatus. In practice it is impossible that Bob has detectors with perfect efficiency. The quantum efficiency $\eta_{D}$ is the probability that the detector registers a count when one photon comes in. And the error rate caused by the detection apparatus is $\varepsilon_{D}$, which the main error source is the dark count d.
\end{enumerate}

To simplify the protocol, assume that Alice’s emission apparatus and Bob’s detection apparatus are both prepared by the trusted third party. They all know the quantum efficiency $\eta_{D}$, the dark count rate $d$ of Bob’s detectors and the typical average photon number $\mu_{S}$ of light pulses that Alice will send to Bob and the transfer efficiency $\eta_{C}$. Suppose the the typical average photon number in detectors is $\mu=\mu_{S} \eta_{C} \eta_{D}$, the overall error rate is $\varepsilon=1-(1-\varepsilon_S)(1-\varepsilon_C)(1-\varepsilon_D)$.

\section{Practical protocols}
\begin{Definition}
\emph{\textbf{(Random Oblivious Transfer (R-OT) Channel)}}
Alice sends a random bit $r$ to Bob via a channel, if
\begin{enumerate}
\item Bob obtains the bit value $r$ with a probability $p$ satisfies $0<\beta < p < \alpha$, $\alpha <\frac{1}{2}$, $\alpha$ and $\beta$ are any two real numbers;
\item Alice cannot know whether Bob has get the value of her bit.
\end{enumerate}
Then, the channel is named a R-OT channel (an extended Rabin's OT channel).
\end{Definition}

To construct a quantum R-OT protocol, we use two non-orthogonal states. There is no measuring apparatus can distinguish between these two states with certainty, only some probabilistic information can be obtained. Let Alice send a sequence of photons in two quantum states $|\Psi_{0}\rangle$, $|\Psi_{1}\rangle$ randomly, where $\langle\Psi_{0}|\Psi_{1}\rangle=\cos\varphi$. Here we choose $\varphi=\frac{\pi}{6}$. Although there are kinds of measurements for Bob, to simplify the protocol to be a practical one, we choose an easy one: Bob measures the coming photons in two basis, $B_{0}=\{|\Psi_{0}\rangle, |\Psi_{0}\rangle^{\perp}\}$ and $B_{1}=\{|\Psi_{1}\rangle, |\Psi_{1}\rangle^{\perp}\}$ randomly. If his measurement results in $|\Psi_{x}\rangle$, he could not distinguish which state was sent by Alice. If his measurement results in $|\Psi_{x\oplus 1}^{\perp}\rangle$, orthogonal to $|\Psi_{x\oplus 1}\rangle$, he know that the initial state could not be $|\Psi_{x\oplus 1}\rangle$ and therefore was $|\Psi_{x}\rangle$. In an ideal world, Alice's probability of sending $|\Psi_{x}\rangle$ is $\frac{1}{2}$ and Bob's probability of detecting a proton in $B_{x\oplus 1}$ is also $\frac{1}{2}$. Therefore, the probability of getting a conclusive result is
\begin{eqnarray}\label{eq:pideal}
p_{ideal}=\frac{1}{2}\times\frac{1}{2}\langle\Psi_{x}|\Psi_{x\oplus 1}^{\perp}\rangle \langle\Psi_{x\oplus 1}^{\perp}|\Psi_{x}\rangle\times2=\frac{1}{2}\sin^{2}\frac{\pi}{6}=\frac{1}{8}
\end{eqnarray}
~\\

\begin{Protocol}
\emph{\textbf{Practical quantum R-OT protocol}}
\begin{enumerate}
\item Alice and Bob decide on a series of security parameters.
The number of photons in a pules with typical average photon number of $\mu_{S}$ obey Poisson distribution.
\begin{eqnarray}\label{eq:1}
{p_n}(\mu_{S}) = \frac{{{e^{ - \mu_{S}}}{\mu_{S} ^n}}}{{n{!}}}
\end{eqnarray}
Form Equation (\ref{eq:1}), the probability of no photons in the pulse is ${p_0}(\mu_{S} ) = {e^{ - \mu_{S} }}$. It is easily to see the Poisson probability of detecting the photons in a pulse with typical average photon number $\mu_{S}$ through a channel with transfer efficiency $\eta_{C}$ by a detector with quantum efficiency $\eta_{D}$ is $1 - {e^{ - \mu}}$. They can set the fraction $a$ which is the probability Alice expects Bob to detect successfully around to $1 - {e^{ - \mu}}$ and set error rate $\varepsilon_{set}$ to $\varepsilon$ or a little bit higher to allow other noises. The parameters are in accord with the equation $H(2\varepsilon_{set})<\frac{1}{2}-(1-e^{-\mu_{S}}-\mu_{S} e^{-\mu_{S}})/2a$ to resist photon number splitting attack\cite{OT92}. They agree on a security parameter $N$ which will be used below. They choose an information reconciliation scheme for about $a N$ bits words with expected error rate $\varepsilon_{set}$. After using these, the error rate will reduce to $\varepsilon'_1$.

\item Alice and Bob perform two tests with their apparatus.

First, they compare the sending time $t_{i}$ with the receiving time $t_{i}'$ for each pulse. Since the distance between them is fixed, they can easily get the traveling time $\theta$, i.e. $\theta=t_{i}'-t_{i}$. This test can not only mark the address of each pulse, but also decrease the error caused by noises and dark counts.

Second, Alice sends sequence of pulses through a quantum channel and tells Bob the bases of the pulses through a classical channel. Bob detects the pulses in the other bases. If and only if he detects the pulses successfully with a probability greater than $a$ and a error rate less than $\varepsilon_{set}$, he agrees to continue the protocol. Otherwise, they take counsel together to adjust the parameter $a$ or $\varepsilon_{set}$.

\item Bob prepares a random qubit string $|\Phi_1\rangle,...,|\Phi_n\rangle$ and sends it to Alice, where $|\Phi_i\rangle\in\{|0\rangle, |1\rangle, |+\rangle, |-\rangle\}$

\item Alice generates random bit string
$({r_1},...,{r_{N}})\in{\{0,1\}}^{N}$. When $r_i=0$, she keeps the ith qubit unchanged and sends it back to Bob; when $r_i=1$, she rotates the state along y axis with $\frac{\pi}{6}$, and sends the qubit back to Bob, that is
\begin{equation}\nonumber
\left\{
\begin{aligned}
&r_i=0,~~ |\Phi_i\rangle \longrightarrow |\Phi_i\rangle,\\
&r_i=1,~~ |\Phi_i\rangle \longrightarrow |\Phi_i+\frac{\pi}{6}\rangle
\end{aligned}
\right.
\end{equation}
She also tells Bob the sending time $t_{i}$ of each pulses through a classical channel, $i\in{\{1,2,...,N\}}$.

\item Bob tells Alice the receiving time of the receiving pulses. The number of them is expected to be $a N$. He chooses ${B_0}$ or ${B_1}$ randomly to measure the pulses coming from Alice, where $|\Psi_0\rangle=|\Phi_i\rangle$ and $|\Psi_1\rangle=|\Phi_i+\frac{\pi}{6}\rangle$. He records the receiving time $t_{i}'$ of each pulse and compares with the sending time. If and only if $t_{i}'=t_{i}+\theta$, he admits $|\Psi_{r_{i}}\rangle$ a receiving pulse. From these receiving pulses, if and only if his measurement results in state $|\Psi_{x}\rangle^{\perp}$, he accepts a pulse as a conclusive pulse and takes the bit value of this pulse as $x\oplus 1$.

\item Alice and Bob use the information reconciliation scheme mentioned in Step 1 to reduce the error in the receiving bits.
\end{enumerate}
\end{Protocol}

If Alice can get a bit's value and ensure that it is a conclusive bit, the qubit Bob obtained must be in a pure state. Therefore, Alice cannot execute EPR attack, and then, she cannot know whether a bit with a given value has been taken as a conclusive bit by Bob.

\begin{Protocol}\textbf{\emph{OT$_{1}^{2}$ protocol}}
\begin{enumerate}
\item Alice and Bob execute protocol 1. Bob's probability of getting a conclusive bit is $p_{con}(\mu)$, which is to be analyzed later. Therefore Bob is supposed to obtain $N {p_{con}(\mu)}$ conclusive bits after this step.
\item Bob selects $I=\{i_{1},\ldots, i_{k}\}$ and $J=\{j_{1},\ldots, j_{k}\}$ with ${I}\cap{J}=\emptyset$. The $k$ bits $r_{i_{1}},\ldots, r_{i_{k}}$ are chosen from his conclusive bits. In practice, if the conclusive bits in Bob's hand are a little less than $k$, he should random adds it to $k$.
\item Bob chooses a random bit with value $m$. If $m=0$, he sends $\{X, Y\}= \{ I, J\}$ to Alice. Otherwise, he sends $\{X, Y\}= \{ J, I\}$.
\item
After receiving $(X, Y)$, Alice generates local random bit-string $R_{0},R_{1}\in \{0,1\}^{k}$ and encrypts her messages $b_{0}$, $b_{1}$ as
\begin{equation}\nonumber
\left\{
\begin{array}{c}
c_{0}=E_{x}(R_{0},b_{0})\in \{0,1\}^{k+1},x=(r_{i}|i\in X),\\
c_{1}=E_{y}(R_{1},b_{1})\in \{0,1\}^{k+1},y=(r_{i}|i\in Y),
\end{array}
\right.
\end{equation}
 where $x,y$ are the encryption keys. Alice sends $c_0$, $c_1$ to Bob and keeps $R_0$, $R_1$ secret.
\item Bob decrypts that coming from Alice to obtain $(R_m, b_{m})=D_{key}(c_{m})$ with $key=(r_{i}|i\in I)$.

\end{enumerate}
\end{Protocol}

In practice, the physical system and the coded bit string are necessary to cause error. Therefore, this kind of OT protocols has a certain probability of error. But it does not impact us to construct a bit commitment protocol.

If Alice regards the two bits in Protocol 2 as a committed bit, her best cheating strategy is to commit $0$ with a probability $\frac{1}{2}$ and commit $1$ with a probability $\frac{1}{2}$. And change half of them in open phase. For each bit, the probability that a cheating Alice can not be detected is $25\%$. Thus even the error rate of OT$_{1}^{2}$ is $20\%$, a cheating Alice could be detected by Bob.

We can construct a bit commitment protocol by executing the protocol $l$ times as follows:
\begin{Protocol}\textbf{\emph{Bit commitment protocol}}
~~~~~~~~~~~~~~~~~~~~~~~~~~~~~~~~~~~~~~~~~~~~~~~~~~~~~~~~~~~~~~~~~~~~~~~~~~~~~~~~~~~~~~~~~~~~~~~~~~~~~~~~~

Commit phase:
\begin{enumerate}
\item Alice randomly divides her commit value as $b= b_{0}^{(i)}\oplus b_{1}^{(i)}$, $i= 1,\ldots, l$. Then there are $l$ same value of $b$.
\item Bob generates local random numbers $\{m_{i}=0, 1| i=1,\ldots, l\}$.
\item  Alice executes protocol 2 with Bob $l$ times, and Bob can obtain the values $\{b_{m_{i}}^{(i)}|i=1,\ldots, l\}$ and $\{R_{m_{i}}^{(i)}|i=1,\ldots, l\}$.
\end{enumerate}
~~~~~Open phase:
\begin{enumerate}
\item Alice opens $\{b_{0}^{(i)}, b_{1}^{(i)}; R_{0}^{(i)}, R_{1}^{(i)}; r_{i_{1}(i)}^{(i)},\ldots,r_{i_{k}(i)}^{(i)}; r_{j_{1}(i)}^{(i)},\ldots,r_{j_{k}(i)}^{(i)}|i= 1,\ldots l\}$.
\item Bob verifies whether $\{b_{0}^{(i)}, b_{1}^{(i)}; R_{0}^{(i)}, R_{1}^{(i)}; r_{i_{1}(i)}^{(i)},\ldots,r_{i_{k}(i)}^{(i)}; r_{j_{1}(i)}^{(i)},\ldots,r_{j_{k}(i)}^{(i)}|i= 1,\ldots l\}$ is consistent with his $\{b_{m_{i}}^{(i)}; R_{m_{i}}^{(i)}; r_{i_{1}(i)}^{(i)},\ldots,r_{i_{k}(i)}^{(i)}|i= 1,\ldots l\}$ and those conclusive bits in J. If the consistency holds, he admits Alice's commit value as $b$.
\end{enumerate}
\end{Protocol}

\section{Analysis on error rate}

Alice and Bob use $(63,57,3)$ Hamming code and the specific information reconciliation scheme executed in Protocol 1 Step 5 is as follows.
\begin{enumerate}
\item $l_{obt}$ denotes the number of $k$ bits in set $I$ or $J$ before using this scheme.
 Alice divides both of $l_{obt}$ bits into 63-bit blocks and performs the wire link permutation W on it.
    When $l_{obt}=63\left\lceil{\frac{l_{obt}}{63}}\right\rceil-\Delta$, $\Delta$ bits of the block in front should be added to the last block. Then calculate the syndromes $s_{A_i}$ and discard the check bits of each block. Repeat above operations four times and send these syndromes to Bob.
\item Bob divides his $l_{obt}$ bits into 63-bit blocks and performs the wire link permutation W on it. When $l_{obt}=63\left\lceil{\frac{l_{obt}}{63}}\right\rceil-\Delta$, $\Delta$ bits of the block in front should be added to the last block. For each round, he calculates the syndromes $s_{B_i}$ and $s_i=s_{A_i}\oplus s_{B_i}$. Correct the error in each block and discards all check bits.
\end{enumerate}
After discard all check bits, the remain bits as the bits in set $I$ or $J$ in Protocol 2.
\begin{eqnarray}
k=57\left\lceil{\frac{l_{obt}}{63}}\right\rceil-\Delta=l_{obt}-6\left\lceil{\frac{l_{obt}}{63}}\right\rceil.
\end{eqnarray}
Suppose the error rate of each bit in Protocol 1 is $\varepsilon_1=0.3\%$. After information reconciliation, the error rate can be reduced to $\varepsilon'_1=0.0757\%$\cite{IR}.

\begin{figure}[H]
\centering
\includegraphics[width=0.6\textwidth]{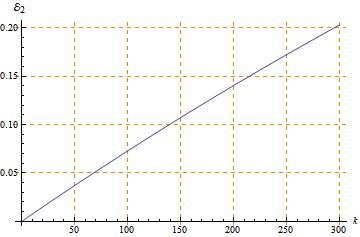}
\caption{The error rate of Protocol 2 with respect to the size of sets}
\label{fig:error2}
\end{figure}

Assume as long as there is one bit error in key used in the decryption algorithm, Bob can not obtain $b_m$ or $R_m$ in Protocol 2. The error rate of Protocol 2 is $\varepsilon_2$. The relation of $\varepsilon_2$ and $\varepsilon'_1$ is
\begin{eqnarray}\label{eq:error}
\varepsilon_2=1-(1-\varepsilon'_1)^k
\end{eqnarray}
When $\varepsilon'_1=0.0757\%$, $\varepsilon_2$ is shown in Figure {\ref{fig:error2}}.

When the sets contain less than $295$ bits, the error rate of Protocol 2 can be less than $20\%$.

\section{Analyze what Bob can obtain and determine the parameteres}

\subsection{Analysis on honest Bob}
Let $|n_{0}\rangle$ and $|n_{\frac{\pi}{6}}\rangle$ denote n-photon state of polarization $0$ and $\frac{\pi}{6}$, respectively. For an honest Bob, if he chooses the measurement bases $B_{1}$ to detect $|1_{0}\rangle$, the probability of the state collapse to $|1_{\frac{2\pi}{3}}\rangle$ is $\frac{1}{4}$. For $|n_{0}\rangle$, the probability of at least one of the photons collapse to the state with polarization of $\frac{2\pi}{3}$ is $1-\left(\frac{3}{4}\right)^{n}$. Therefore, the probability of getting a conclusive resulting in a pulse which contains $n$ photons is
\begin{eqnarray}\label{eq:p(n)}
p(n)=\frac{1}{2}\times\left[1-\left(\frac{3}{4}\right)^{n}\right].
\end{eqnarray}
From Equation(\ref{eq:1}) and Equation(\ref{eq:p(n)}), the probability of getting a conclusive bit in a pulse is
\begin{eqnarray}\label{eq:pcon}
\begin{aligned}
p_{con}(\mu)&=\sum\limits_{n=1}[p_n(\mu)\times{p(n)}]\\
&=\sum\limits_{n=1} \left\{\frac{1}{2}\left[1-\left(\frac{3}{4}\right)^{n}\right]\frac{e^{-\mu} \mu^n}{n!}\right\} \\
&=\frac{1}{2}\left[\sum\limits_{n=1}\frac{e^{-\mu} \mu^n}{n!}-e^{-\frac{\mu}{4}}\sum\limits_{n=1}\frac{{e^{-\frac{3\mu}{4}} \left({\frac{3\mu}{4}}\right)^n}}{n!}\right]\\
&=\frac{1}{2}\left(1-e^{-\frac{\mu}{4}}\right).
\end{aligned}
\end{eqnarray}

\begin{figure}[H]
\centering
\includegraphics[width=0.6\textwidth]{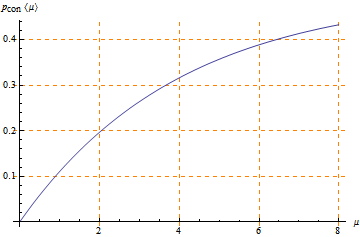}
\caption{The probability of an honest Bob gets a conclusive bit with respect to $\mu$}
\label{fig:honestp}
\end{figure}

It can be seen that an honest Bob is supposed to obtain $N {p_{con}(\mu)}$ conclusive bits. The probability of getting a conclusive bit in one pulse with different $\mu$ can be seen in Figure {\ref{fig:honestp}}. The larger $\mu_S$ of emission apparatus and more efficient detector they use, the higher efficiency the protocol is.

\subsection{Analysis on malicious Bob}
A malicious Bob can separate $n$ photons by photon number splitting attack. For single-photon, the successful probability of optimal measurement to differentiate the two non-orthogonal states is $1-cos\varphi$, which has been proven\cite{Ivanovic,Peres88,LIGE}. For $n$ photons, a malicious Bob's probability of differentiating the two-orthogonal sates is

\begin{eqnarray}\label{eq:diff}
p'(n)=1-cos^n\varphi.
\end{eqnarray}
Then a malicious Bob using photon number splitting attack and optimal measurement for single-photon can get conclusive bits with the probability of

\begin{eqnarray}\label{eq:p'con}
p'_{con}(\mu)=\sum\limits_{n=1}p_n(\mu)\times{p'(n)}=1-e^{-\mu(1-\frac{\sqrt{3}}{2})}.
\end{eqnarray}

Here we consider that the malicious Bob has an ideal detector, the quantum efficient $\eta'_D$ of which is $1$. Thus, $\mu'=\mu_{S} \eta_C=\frac{\mu}{\eta_{D}}$. Assume the protocols are executed over atmospheric channel, the quantum efficiency of honest Bob's detector $\eta_D$ is $80\%$ and this kind of detector has been already realized in laboratory. The cheating Bob's probability of getting a conclusive bit is
\begin{eqnarray}\label{eq:greater}
p''_{con}(\mu_S)=1-e^{-\frac{5\mu}{4}(1-\frac{\sqrt{3}}{2})}
\end{eqnarray}
A malicious Bob will get about $1-e^{-\frac{5\mu}{4}(1-\frac{\sqrt{3}}{2})}N$ conclusive bits.
\begin{figure}[H]
\centering
\includegraphics[width=0.6\textwidth]{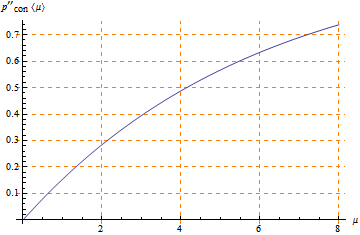}
\caption{The probability of a malicious Bob gets a conclusive bit with respect to $\mu$}
\label{fig:cheatingp}
\end{figure}

\subsection{Contrastive analysis}
\begin{figure}[H]
\centering
\includegraphics[width=0.6\textwidth]{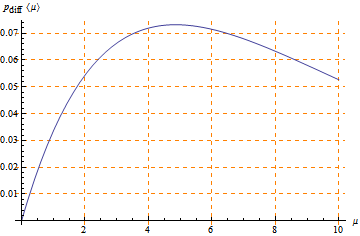}
\caption{The difference between an honest Bob's probability of obtaining a conclusive bit and half of a malicious Bob's probability of obtaining a conclusive bit with respect to $\mu$}
\label{fig:u}
\end{figure}

The difference between an honest Bob's probability of obtaining a conclusive bit and half of a malicious Bob's probability of obtaining a conclusive bit is $p_{diff}(\mu)=p_{con}(\mu)-\frac{1}{2}p''_{con}(\mu)$, which can be seen in Fig {\ref{fig:u}}.

When $\mu=4.85$, the function has a maximum $0.0732$.
The probability of obtaining $i$ conclusive bits is $p_{obt}$, which is referred to the binomial distribution.

\begin{figure}[H]
\centering
\includegraphics[width=0.6\textwidth]{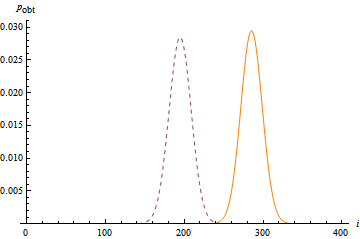}
\caption{The solid line denotes the probability of an honest Bob obtains $i$ conclusive bits when $N=800$, $\mu=5$. It can be seen an honest Bob can obtain more than $259$ conclusive bits with a great probability. The dashed line denotes the probability of a malicious Bob obtains $i+259$ conclusive bits.}
\label{fig:pobt}
\end{figure}

\subsection{Determine the parameters in practical protocols}
Suppose the probability of the cases that the number of conclusive bits obtained by an honest Bob is equal to or less than $l_{obt}$ be $p_{1}$, and the probability of the cases that the number of conclusive bits obtained by a malicious Bob is equal to or greater than $2 l_{obt}$ be $p_{2}$.
\begin{eqnarray}\label{eq:p}
p_1=\sum\limits_{i=0}^{l_{obt}} C_N^i[p_{con}(\mu)]^i [1-p_{con}(\mu)]^{N-i}\\
p_2=\sum\limits_{i=2 l_{obt}}^{N} C_N^i[p'_{con}(\mu)]^i [1-p'_{con}(\mu)]^{N-i}
\end{eqnarray}
As the honest Bob should execute Protocol 2 successfully and the malicious Bob can not obtain both $b_0$ and $b_1$, $p_{1}$ and $p_{2}$ should be small enough.

The probability of an honest Bob cannot execute Protocol 2 successfully is $p$.
\begin{eqnarray}
p=1-(1-\varepsilon_2)(1-p_1).
\end{eqnarray}
To detect a cheating Alice, $p$ should less than $20\%$. Given a $\varepsilon_2$, $p_1$ has a upper bound $p_{1t}$ to ensure $p\leq20\%$.
To ensure the concealing of the bit commitment protocol, $p_2$ should be controlled to be a magnitude of $10^{-6}$.

\begin{table}[htbp]
\caption{When $p=20\%$, $N=800$, $p_2$ is controlled to be a magnitude of $10^{-6}$, the selection of parameters with different $\mu$.}
\begin{tabular}{lclclclclclclclclcl}
\toprule
\multicolumn{1}{c}{$\mu$} &
\multicolumn{1}{c}{$p_{con}(\mu)$} &
\multicolumn{1}{c}{$p'_{con}(\mu)$} &
\multicolumn{1}{c}{$l_{obt}$} &
\multicolumn{1}{c}{$k$} &
\multicolumn{1}{c}{$\varepsilon_2$} &
\multicolumn{1}{c}{$p_{1t}$} &
\multicolumn{1}{c}{$p_1$} &
\multicolumn{1}{c}{$p_2$}                \\
\midrule
$2$ & $0.197$ & $0.285$ & $143$ & $131$ & $0.0944$ & $0.117$ & $0.107$ & $3.25\times10^{-6}$  \\

$3$ & $0.264$ & $0.395$ & $190$ & $172$ & $0.122$ & $0.0887$ & $0.0484$ & $1.85\times10^{-6}$  \\

$4$ & $0.316$ & $0.488$ & $228$ & $210$ & $0.147$ & $0.0621$ & $0.0312$ & $1.53\times10^{-6}$  \\

$5$ & $0.357$ & $0.567$ & $260$ & $236$ & $0.164$ & $0.0435$ & $0.0324$ & $7.45\times10^{-7}$  \\

$6$ & $0.388$ & $0.634$ & $283$ & $259$ & $0.178$ & $0.0267$ & $0.0236$ & $4.73\times10^{-6}$  \\

\bottomrule
\end{tabular}
\end{table}

When $\mu$ is too low, the difference between the probability of obtaining a conclusive bit by an honest or a malicious Bob is not large enough to select proper parameters. When $\mu$ is too large, a proper $k$ is too large to lead to a large $\varepsilon_2$ and we cannot select proper parameters either.
It can be seen from Table $1$, when $2\leq\mu\leq6$, we can always find other proper parameters to execute the protocols successfully.

\section{Security}
\subsection{Concealing of bit commitment protocol}
Suppose the probability of a malicious Bob obtains both of two messages in $OT^2_1$ protocol $p_2$ is controlled to be a magnitude of $10^{-6}$, the times of executing $OT^2_1$ protocol in bit commitment protocol $l$ is $25$, a malicious Bob can obtain what Alice has committed before open phase with a probability of
\begin{eqnarray}
p_{br}=1-(1-10^{-6})^{25}\approx2.5\times10^{-5}
\end{eqnarray}
In practical protocol, the probability of breaking the concealing of bit commitment around $5\times10^{-5}$ is allowed.

\subsection{Binding of bit commitment protocol}

When Alice tries to attack the binding of bit commitment protocol, she changes the value of $b_{0}^{(i)}$ or $b_{1}^{(i)}$. But some of these values are fixed after execute Protocol 2, and Alice has no idea about which bits Bob obtains, even no-go attack can not help. If $OT^2_1$ is secure, what she only can do is to choose randomly the changing bit between $b_{0}^{(i)}$ and $b_{1}^{(i)}$ for each $i$. The probability that Bob can obtain a correct $b_{0}$ or $b_{1}$ successfully is $1-p=0.8$ and the probability that a cheating Alice can be detected is $0.4$. Therefore, for $l=25$, Alice's success probability of attacking is
\begin{eqnarray}
p'_{br}=(1-0.4)^{25}\approx2.8\times10^{-6}
\end{eqnarray}
In practical protocol, the probability of breaking the binding of the bit commitment around $2.8\times10^{-6}$ is allowed.

\section{Discussions}
In this paper, we analyze the situation that the protocols are executed on atmospheric window with a high efficiency detector of $80\%$. Otherwise, if a malicious Bob has a greater ability to obtain information near Alice's site and has a super channel, the transfer efficiency could be $100\%$. To defend the attack, the efficiency of transfer and an honest Bob's detector $\eta_C \eta_D$ should be increased to $80\%$.

If we execute the protocols in optical fiber, the bit commitment protocol can be realized between two parties with a long distance. For a malicious Bob who uses photon number splitting attack and has a detector with a efficiency less than $\eta_D/80\%$, the analysis and security of the protocol also hold true. It means that our protocols is probably applied over a long distance in future.

\section{Conclusion}
Based on two non-orthogonal states, we construct a random oblivious transfer protocol. Then we construct a one-out-of-two oblivious transfer protocol on top of the random OT protocol. Finally, we present a bit commitment protocol based on the one-out-of-two protocol. The security of concealing is kept by measurement hypothesis and superposition principle of state in quantum mechanics. Since the no-go theorem type attack can not work because of random $|\Phi_i\rangle$, the binding of the bit commitment protocol is secure\cite{Yang13}. By using weak coherent pulses and allowing some error, our protocols can be applied in practice. With the advent of the higher efficiency detectors in optical fiber, our protocol can be realized with a long distance.

~~~~~~~~~~~~~~~~~~~~~

%\softraggedright
%\itemsep=-4pt plus.2pt minus.2pt  %% sets the vertical space between items
%\small

\bibliographystyle{elsarticle-num-names}
\bibliography{Endnote}

\end{document}